\documentclass[12pt]{article}
\usepackage[dvips]{graphicx}
\usepackage{pdproc}
\usepackage{url}

  \makeatletter 
  \def\@cite#1{[#1]} 
  \makeatother    
\begin{document}

\renewcommand{\thefootnote}{\alph{footnote}}

\title{
Neutralino polarization effect in the squark cascade decay at LHC%
\footnote{
Talk based on the work \cite{Goto:2004cp}, presented at the 12th
International Conference on Suypersymmetry and Unification of the
Fundamental Interactions (SUSY04), June 17-23, 2004, in Epochal Tsukuba,
Tsukuba, Japan.
}
}

\author{ TORU GOTO }

\address{ 
YITP, Kyoto University,
Kyoto, 606-8502 Japan
\\ {\rm E-mail: goto@yukawa.kyoto-u.ac.jp}
}

\abstract{
We study the effect of the neutralino polarization in a squark cascade
decay $\tilde{q} \to q \tilde{\chi}^0_2 \to q l^\pm \tilde{l}^\mp \to
 q l^\pm \tilde{l}^\mp \to q l^\pm l^\mp \tilde{\chi}^0_1$.
Charge asymmetry in the lepton-jet invariant mass distribution appears
depending on the chirality structure of the sfermion-fermion-neutralino
coupling.
With use of the Monte Carlo simulation, we show that the asymmetry is
measurable in LHC.
We also show that the distribution of the charge asymmetry is sensitive
to the (s)lepton flavor.
}

\normalsize\baselineskip=15pt

\section{Introduction}

One of the important objectives in the LHC experiments is to obtain the
information about the interactions of the SUSY particles.
In this work \cite{Goto:2004cp}, we study the following squark cascade
decay process
\begin{equation}
\tilde{q}\to q \tilde{\chi}^0_2 \to q l^\pm_1 \tilde{l}^\mp \to
q l^\pm_1 l^\mp_2 \tilde{\chi}^0_1,
\label{eq:decay}
\end{equation}
and investigate if SUSY study at the LHC can provide an information on
the chiral nature of the sfermions.

Due to the chirality structure of the squark-quark-neutralino coupling,
the $\tilde{\chi}^0_2$ is polarized.
The polarization of $\tilde{\chi}^0_2$ then affects the angular
distribution of the slepton in the neutralino decay.
The polarization dependence of the angular distribution eventually shows
up in the charge asymmetry in the $m(q l^\pm_1)$ distribution, because
the polarization dependent part of the amplitude flips under the charge
conjugation transformation \cite{Richardson:2001df}.

We study the charge asymmetry in the minimal supergravity (mSUGRA)
model, taking account of the slepton left-right mixing effect.
We show that the charge asymmetry, as well as the
$\tilde{\chi}^0_2\to l\tilde{l}$ branching ratio, depends on the lepton
flavor $l=e$, $\mu$ or $\tau$.
We perform a Monte Carlo simulation to demonstrate the detectability at
the LHC.

\section{Decay Distribution and Charge Asymmetry}

The sfermion-fermion-neutralino interaction Lagrangian is written as
\begin{eqnarray}
  {\cal L} &=& -\frac{ g_2 }{ \sqrt{2} }
  \sum_{i,\alpha}
  \bar{\tilde{\chi}}^0_i
  \left(
      L_{i\alpha}^{f}\frac{1-\gamma_5}{2}
    + R_{i\alpha}^{f}\frac{1+\gamma_5}{2}
  \right)
  f \tilde{f}^*_\alpha
  + {\rm H.c.},
\label{eq:Lag}
\end{eqnarray}
where $f=q,l$ ($\tilde{f}=\tilde{q},\tilde{l}$) refers to quark (squark)
and lepton (slepton) fields.
$i=1,2,3,4$ and $\alpha=1,2$ are the suffices for the mass
eigenstates of the neutralinos and the sfermions, respectively.
$g_2$ is the SU(2) gauge coupling constant.
The coefficients $L_{i\alpha}^{f}$ and $R_{i\alpha}^{f}$ are determined
by the SU(2)$\times$U(1) gauge couplings, the Yukawa couplings, the
neutralino mixing matrix and the sfermion left-right mixing angles.

The angular distribution of the decay chain (\ref{eq:decay}) is given as
\begin{eqnarray}
  \frac{ d^3\Gamma }{ d\cos\theta_{\tilde{l}}\,
    d\cos\theta_{\tilde{\chi}^0_1}\, d\phi_{\tilde{\chi}^0_1} }
  &=&
  \frac{1}{8\pi} \Gamma(\tilde{q}_\beta \to q \tilde{\chi}^0_2 )
  Br(\tilde{\chi}^0_2 \to l^\pm_1 \tilde{l}^\mp_\alpha )
  Br(\tilde{l}^\mp_\alpha \to l^\mp_2 \tilde{\chi}^0_1 )
\nonumber\\&&\times
  \left[
    1 \mp A(l)\cos\theta_{\tilde{l}}
  \right],
\label{eq:dist}
\\
  A(l) &=&
  \frac{ |L_{2\beta}^{q}|^2 - |R_{2\beta}^{q}|^2 }
       { |L_{2\beta}^{q}|^2 + |R_{2\beta}^{q}|^2 }
  \cdot
  \frac{ |L_{2\alpha}^{l}|^2 - |R_{2\alpha}^{l}|^2 }
       { |L_{2\alpha}^{l}|^2 + |R_{2\alpha}^{l}|^2 },
\label{eq:asym}
\end{eqnarray}
where $\Gamma$ and $Br$ denotes the decay width and branching ratio,
respectively.
$\theta_{\tilde{l}}$ is the angle between the momenta of the quark and
the lepton $l_1$ in the $\tilde{\chi}^0_2$ rest frame,
$\theta_{\tilde{\chi}^0_1}$ is the angle between the two lepton momenta
in the slepton rest frame, and $\phi_{\tilde{\chi}^0_1}$ is the angle
between the decay planes of $\tilde{q} \to q l^\pm_1 \tilde{l}^\mp$ and
$\tilde{\chi}^0_2 \to l^\pm_1 l^\mp_2 \tilde{\chi}^0_1$.
Since squark and slepton decays are spherically symmetric in the rest
frames of the decaying particles, the angular distribution is flat over
$\cos\theta_{\tilde{\chi}^0_1}$ and $\phi_{\tilde{\chi}^0_1}$.
The $\theta_{\tilde{l}}$ dependence, which comes from the polarization of
$\tilde{\chi}^0_2$, eventually shows up in the quark-lepton invariant
mass $m(ql)$ distribution.
Since $l_1$ from the neutralino decay and $l_2$ from the slepton decay
are indistinguishable, we study the charge asymmetry between the
$m(ql^\pm)$ distributions taking both $l_1$ and $l_2$ into account
\cite{Barr:2004ze}.
The $m(ql^+)$ distribution consists of $l_1^+$ from
$\tilde{\chi}^0_2 \to l_1^+\tilde{l}^-$
and $l_2^+$ from
$\tilde{\chi}^0_2 \to l_1^-\tilde{l}^+ \to l_1^-l_2^+\tilde{\chi}^0_1$.

We consider a ``typical'' case of the mSUGRA, where the wino component
dominates $\tilde{\chi}^0_2$ and the bino component dominates
$\tilde{\chi}^0_1$.
We can safely neglect the left-right mixing of the squarks because the
process we consider is the decay of the first generation squark.
The first decay process $\tilde{q} \to q \tilde{\chi}^0_2$ in
Eq.~(\ref{eq:decay}) occurs predominantly through the
$\tilde{q}_L$-$q$-$\tilde{W}$ coupling so that the first factor of the
right-hand side of Eq.~(\ref{eq:asym}) is very close to unity.

As for the slepton mass matrix, the right-handed slepton mass parameter
becomes smaller than the left-handed one due to the running effect.
Therefore the lighter slepton, $\tilde{l}_1$, is $\tilde{l}_R$-like in
the most of the parameter space.
In the $\tilde{l}_1$-$l$-$\tilde{\chi}^0_2$ couplings,
the main component of $R_{21}^{l}$ comes from the U(1) gauge coupling
and is approximately universal for the lepton flavor.
On the other hand, the SU(2)$\times$U(1) gauge coupling term in
$L_{21}^{l}$ appears only through the left-right mixing of the sleptons,
which is proportional to the lepton mass, so that $L_{21}^{l}$ depends
on the lepton flavor.

For the (s)electron, $L_{21}^{e}$ is negligibly small compared to
$R_{21}^{e}$ and the charge asymmetry is maximal ($A(e)\approx -1$).
The effect of the left-right mixing and the Yukawa coupling is quite
significant for the (s)tau mode.
The mixing angle of the stau is typically $O(1)$ for large $\tan\beta$.
Then $L_{21}^{\tau}$ dominate over $R_{21}^{\tau}$ and the behavior of
the charge asymmetry is opposite to the electron case.
The left-right mixing effect may be observed even in the (s)muon case,
since $L_{21}^{\mu}$ is enhanced by $O(m_\mu/m_e)$ compared to
$L_{21}^{e}$.
For a relatively large $\tan\beta$, it is possible that $L_{21}^{\mu}$
and $R_{21}^{\mu}$ are of the same order of magnitude.

\section{Numerical Results}

We perform a Monte Carlo simulation for mSUGRA benchmark points SPS1a
and SPS3 \cite{Allanach:2002nj}.
SPS1a is given by $m_0=100$~GeV, $M_{1/2}=250$~GeV, $A_0=-100$~GeV,
$\tan\beta=10$ and $\mu>0$%
\footnote{
$m_0$, $M_{1/2}$ and $A_0$ are the common scalar mass, the gaugino mass and
the trilinear scalar coupling at the GUT scale, respectively,
$\tan\beta=\langle h_2\rangle/\langle h_1\rangle$ and $\mu$ is the
Higgsino mass parameter.
}.
Here $m(\tilde{l}_2) > m(\tilde{\chi}^0_2) > m(\tilde{l}_1)$
so that only $\tilde{l}_1$ contributes to the decay chain.
We also show the results for the point with $\tan\beta=20$ and other
parameters are the same as those of SPS1a.
The $\tilde{\mu}$ left-right mixing effect becomes significant for this
point.
Parameters for SPS3 are
$m_0=90$~GeV, $M_{1/2}=400$~GeV, $A_0=0$, $\tan\beta=10$ and $\mu>0$.
In this case, $m(\tilde{\chi}^0_2) > m(\tilde{l}_2) > m(\tilde{l}_1)$
and $\tilde{\chi}^0_2\to \tilde{l}_2$ dominates over $\tilde{l}_1$
process in the kinematically allowed region because the SU(2) gauge
coupling component in $L_{22}^{l}$ dominates over other couplings.

We generated $3\times 10^6$ events for SPS1a and SPS3.
This corresponds to $\int{\cal L}dt=58~{\rm fb}^{-1}$ and
$600~{\rm fb}^{-1}$ for SPS1a and SPS3, respectively.
The mass spectrum, couplings and branching ratios are calculated 
by ISAJET \cite{Baer:1999sp} and interfaced to HERWIG
\cite{Corcella:2000bw} by using ISAWIG program \cite{isawig}.
The events are studied using the fast detector simulator ATLFAST
\cite{atlfast}.
See Ref.~\cite{Goto:2004cp} for detail of the simulation analysis.

\begin{figure}[htb]
\centering
\includegraphics[scale=0.25,clip]{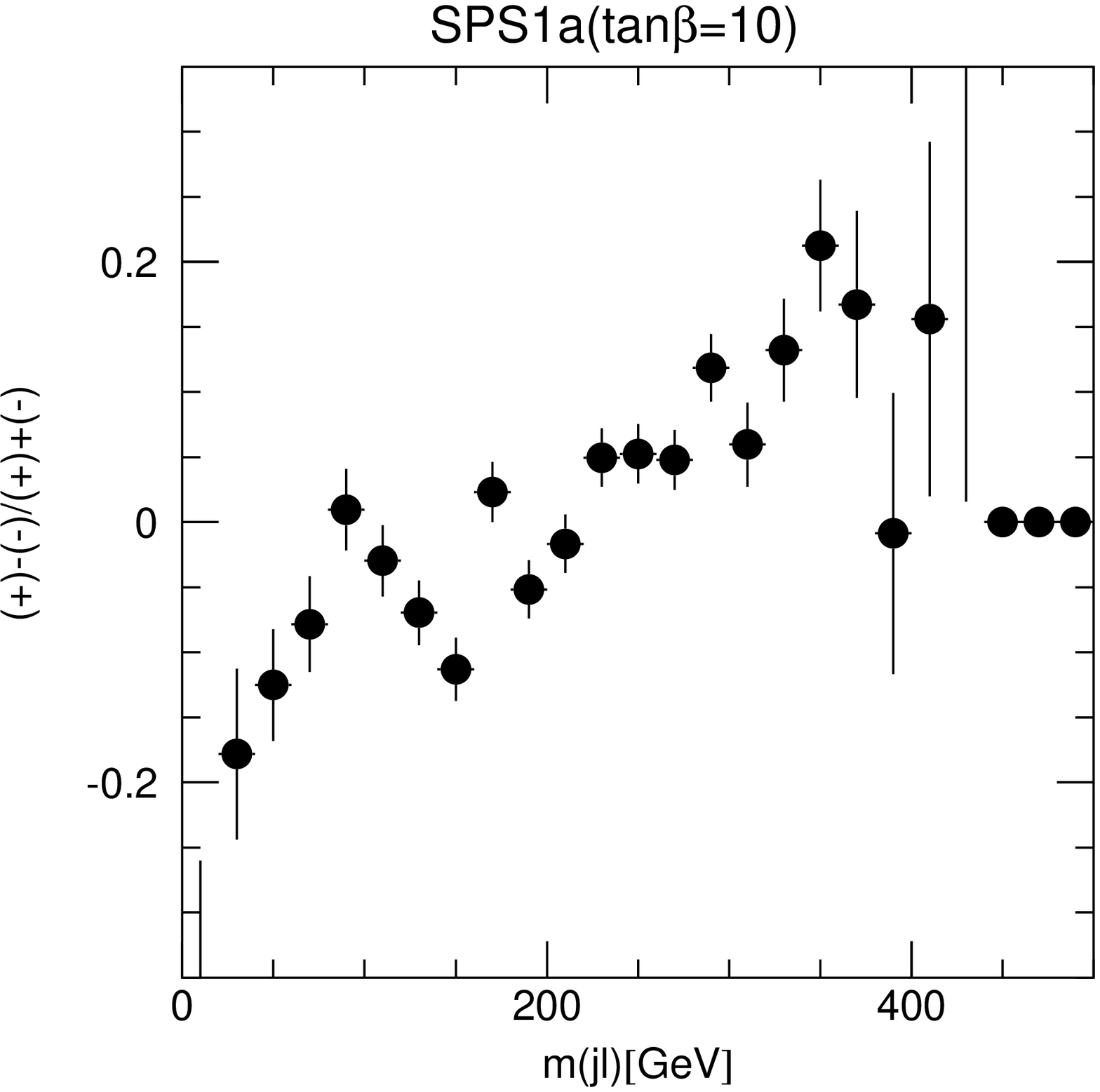}
\makebox[0em][r]{(a)}
\includegraphics[scale=0.27,clip]{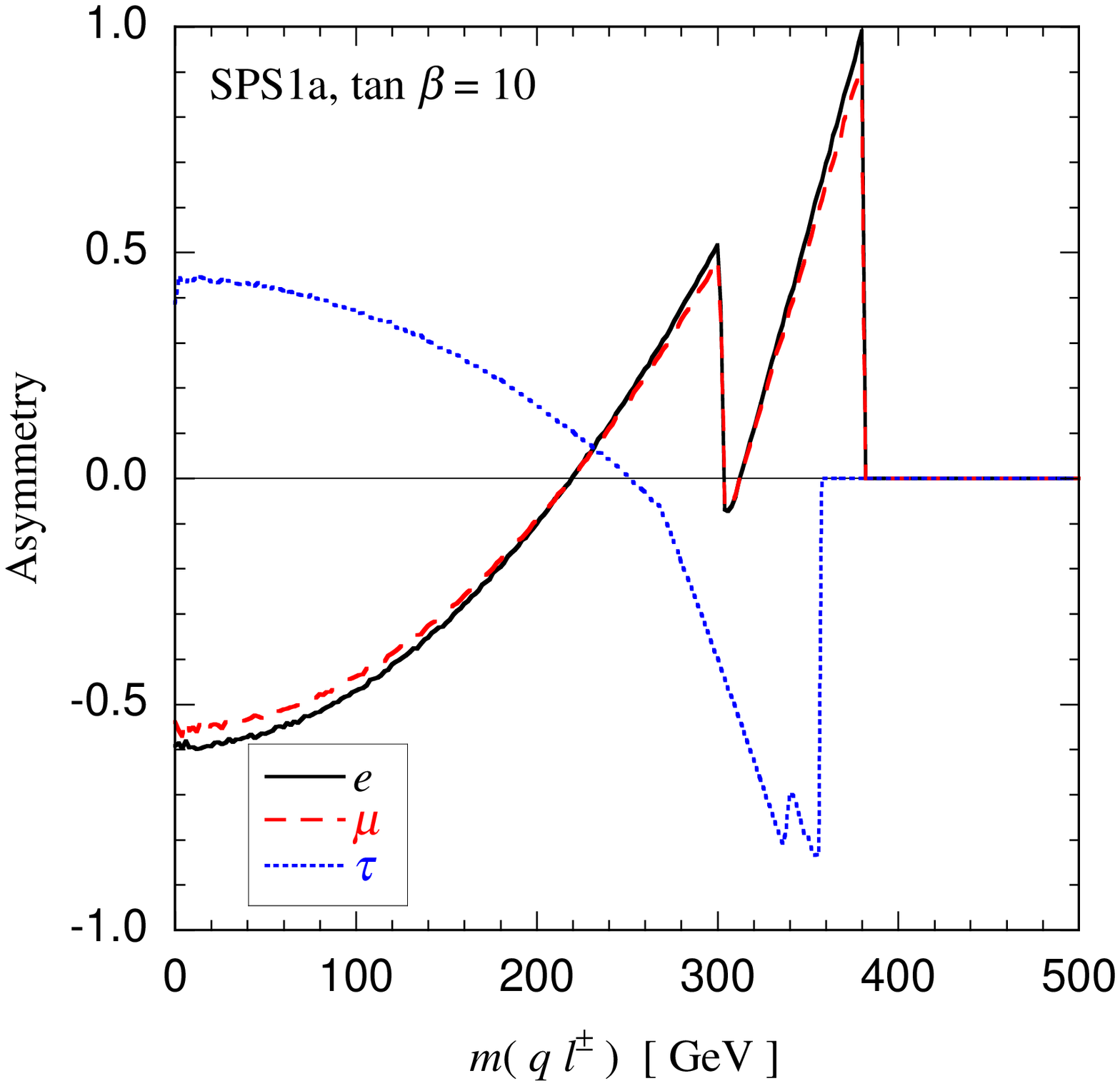}
\makebox[0em][r]{(b)}
\includegraphics[scale=0.27,clip]{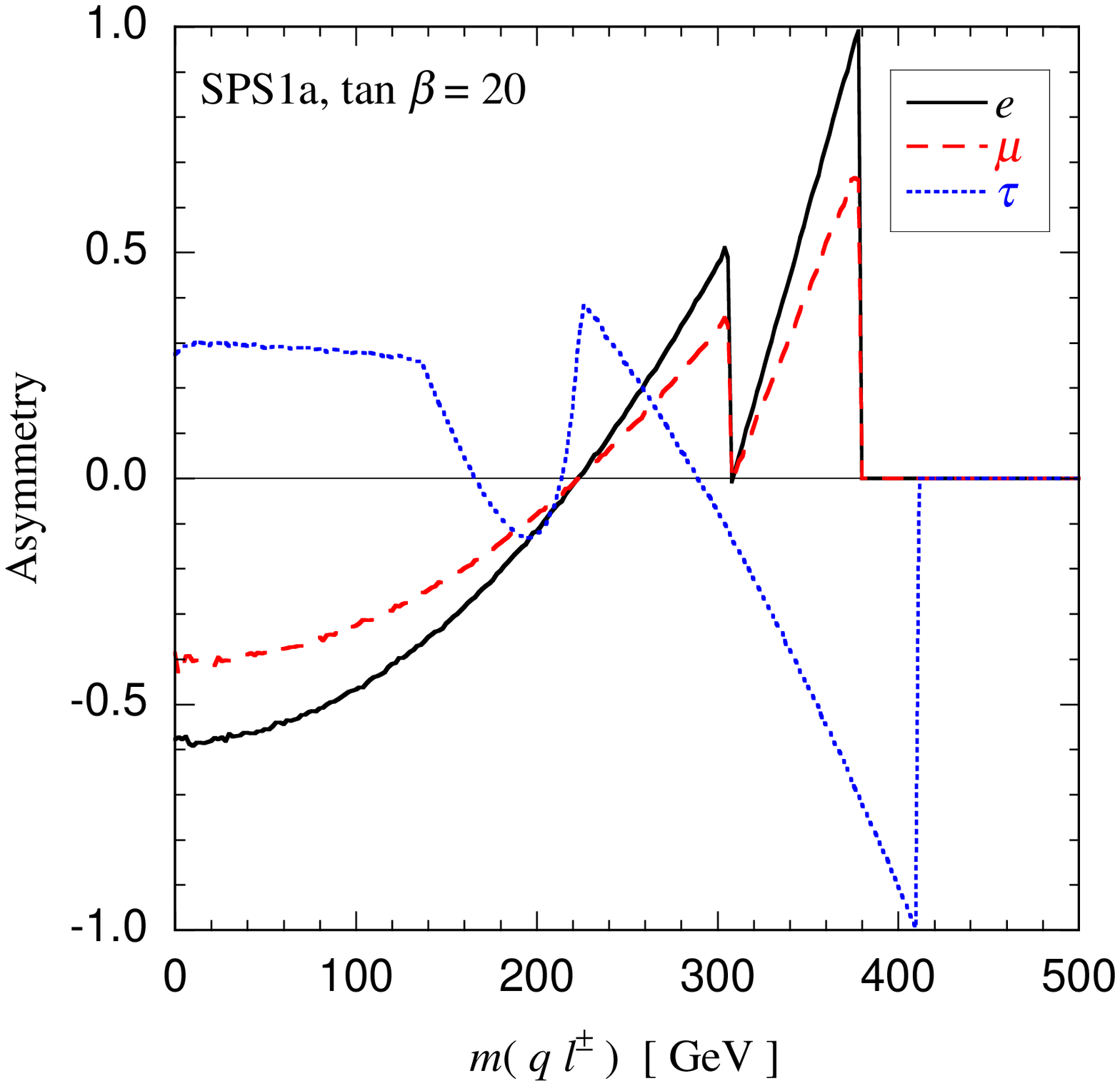}
\makebox[0em][r]{(c)}
\caption{
(a) The reconstructed charge asymmetry for SPS1a.
(b) The calculated charge asymmetry for SPS1a.
(c) The calculated charge asymmetry for SPS1a with $\tan\beta=20$.
}
\label{fig:sps1}
\end{figure}
Fig.~\ref{fig:sps1} shows the charge asymmetry defined by
$[N(ql^+)-N(ql^-)]/[N(ql^+)+N(ql^-)]$, for SPS1a.
(a) is the plot of the simulation data, where $e$ and $\mu$ events are
combined.
Since ISAJET neglect Yukawa couplings for $e$ and $\mu$, the $e$-$\mu$
non-universality discussed in the previous section is not taken into
account in the event generations.
(b) and (c) are calculated theoretical values for $\tan\beta=10$ and
$20$, respectively.
We see that (a) and (b) are qualitatively similar:
the distribution shows negative asymmetry for small $m(jl)$ region and
positive asymmetry near $m(jl)$ endpoint.
The main source of the discrepancy in the magnitudes of the asymmetries
between (a) and (b) is understood as the dilution due to the anti-squark
events.
The dilution factor
$[N(\tilde{q}) - N(\tilde{q}^*)]/[N(\tilde{q}) + N(\tilde{q}^*)]$
is evaluated as 50\% and 58\% for SPS1a and SPS3, respectively, in the
present simulations.
In (b) and (c), we see the lepton flavor dependence of the charge
asymmetry discussed in the previous section.
The asymmetry for $\tau$ mode is opposite to that for $e$ and $\mu$ for
both cases, and $e$-$\mu$ difference is as large as 30\% for
$\tan\beta=20$ case (c).
\begin{figure}[htb]
\centering
\includegraphics[scale=0.25,clip]{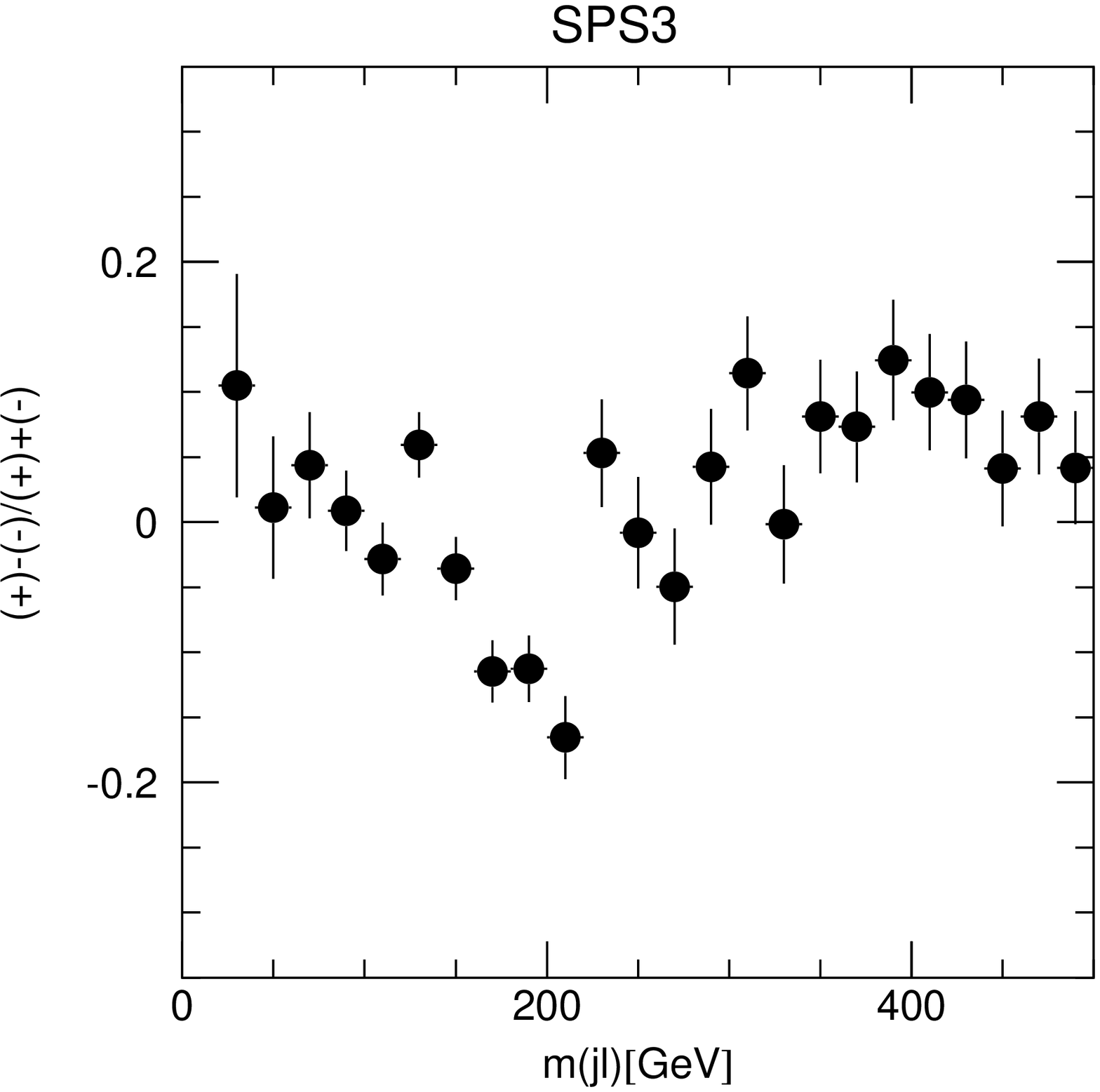}
\makebox[0em][r]{(a)}
\includegraphics[scale=0.27,clip]{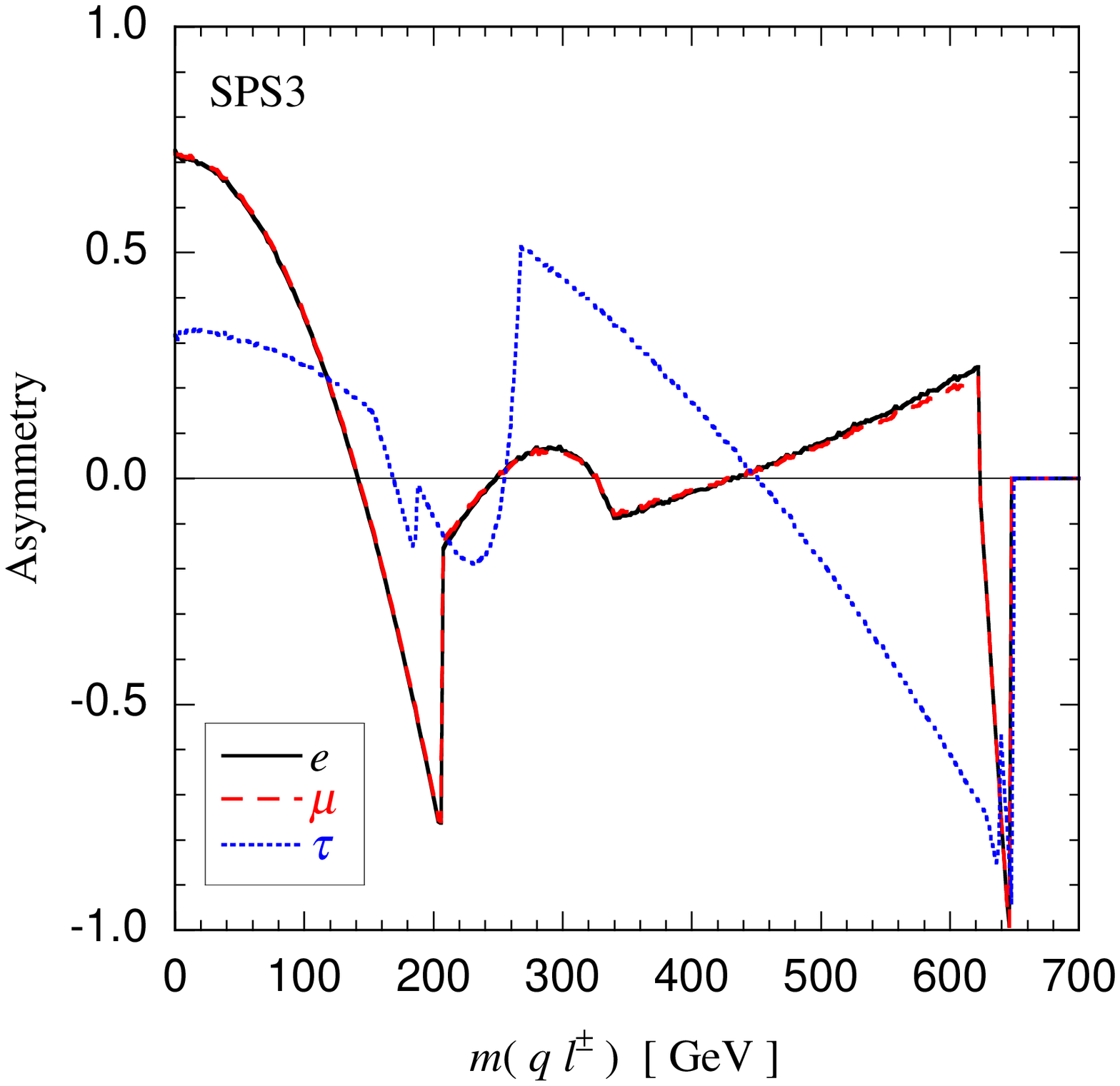}
\makebox[0em][r]{(b)}
\caption{
(a) The reconstructed charge asymmetry and
(b) The calculated charge asymmetry for SPS3.
}
\label{fig:sps3}
\end{figure}
Charge asymmetry for SPS3 is shown in Fig.~\ref{fig:sps3}.
In the $m(ql)<200$~GeV region, the behavior of the asymmetry is similar
for all the lepton flavor, since the lepton mainly comes from the
$\tilde{\chi}^0_2\to l_1\tilde{l}_2$ decay.

Also the branching ratio of $\tilde{\chi}^0_2\to l \tilde{l}_1$ depends
on the lepton flavor.
The decay width for $\tau \tilde{\tau}_1$ mode is much larger than those
for $e \tilde{e}_1$ and $\mu \tilde{\mu}_1$, and
$\Gamma( \tilde{\chi}^0_2 \to \mu \tilde{\mu}_1) >
\Gamma( \tilde{\chi}^0_2 \to e \tilde{e}_1)$ is possible for large
$\tan\beta$.

\section{Conclusion}

In this work, we study the charge asymmetry of $m(jl^\pm)$
distribution in the cascade decay
$\tilde{q}\to q\tilde{\chi}^0_2\to ql\tilde{l}\to qll\tilde{\chi}^0_1$,
in order to study the LHC capability of providing an information on the
chirality structure of the sfermion-fermion-neutralino interaction.
We find that the charge asymmetry is significant for two representative
mSUGRA points, SPS1a and SPS3.
Taking the left-right mixing of the sleptons into account, we show that
the charge asymmetry and the branching ratio of
$\tilde{\chi}^0_2\to l\tilde{l}_1$ is flavor non-universal and that the
LHC can detect $e$-$\mu$ non-universality at
SPS1a($\tan\beta=10$-$20$) for $\int dt {\cal L}=300$~fb$^{-1}$ 
if detection efficiencies of $e$ and $\mu$ 
are understood at the LHC.

\section{Acknowledgment}

The author would like to thank K.~Kawagoe, M.~M.~Nojiri and G.~Polesello.
This work is supported in part by the Grant-in-Aid for the 21st Century
COE ``Center for Diversity and Universality in Physics'' form the MEXT
of Japan.

\bibliographystyle{plain}

\end{document}